\documentclass[prl,twocolumn,showpacs]{revtex4}

\usepackage{amsfonts}
\usepackage{amssymb}
\usepackage{amsbsy}
\usepackage{bm}

\begin{document}

\title{Simple trace criterion for classification of multilayers}
\author{L. L. S\'anchez-Soto, J. J. Monz\'on, T. Yonte}
\address{Departamento de \'{O}ptica, 
Facultad de Ciencias F\'{\i}sicas, 
Universidad Complutense, 28040 Madrid, Spain}
\email{lsanchez@eucmax.sim.ucm.es}

\author{J. F. Cari\~{n}ena}
\address{Departamento de F\'{\i}sica Te\'orica, 
Facultad de Ciencias, Universidad de Zaragoza, 
50009 Zaragoza, Spain}

\date{\today}

\begin{abstract}
The action of any lossless multilayer is described by 
a transfer matrix that can be factorized in terms of
three basic matrices. We  introduce a simple trace 
criterion that classifies multilayers in three classes
with properties closely related with one (and only one)
of these three basic matrices.
\end{abstract}

\maketitle

Any linear system with two input and two output 
channels can be described in terms of a $2 \times 2$
transfer matrix~\cite{YE88,LE87}. In many instances
(e.g., in polarization optics~\cite{AZ87}), we are
interested in the transformation properties of 
quotients of variables rather than on the variables
themselves. Then, the transfer matrix induces a
bilinear transformation that represents the action
of the system in the complex plane~\cite{HA96}.

For layered media the use of $2 \times 2$ matrix  
methods is standard, but the corresponding 
complex-plane representation is not a common tool. 
In fact, it has been recently established that for a 
lossless multilayer the transfer matrix is an element 
of the group SU(1,1)~\cite{MO99}, which is the 
key for a deeper understanding of its behavior and 
simplifies exceedingly its description in terms 
of bilinear transformations.

Moreover,  as we have pointed out recently~\cite{MO01b}, 
the Iwasawa decomposition provides a remarkable 
factorization of the transfer matrix representing 
any multilayer (no matter how complicated it could be)  
as  the product of three matrices of simple interpretation. 
At the geometrical level, such a  decomposition translates 
directly into three actions that  are the basic bricks from 
which any multilayer action is built.

In this Letter we go one step further and show that 
the trace of the transfer matrix allows for a 
classification of multilayers in three disjoint classes 
with properties very close to those appearing in the
Iwasawa decomposition.

To maintain the discussion as self contained as possible 
we briefly summarize the essential  ingredients of 
multilayer optics we shall need  for our purposes. 
The  configuration we consider is a stratified structure 
that consists of a stack of  plane-parallel layers sandwiched  
between two semi-infinite ambient ($a$) and substrate 
($s$) media, which we shall assume to be identical, 
since this is the common experimental case. Hereafter 
all the media are supposed to be lossless, homogeneous, 
and isotropic. 

We assume an incident monochromatic, linearly 
polarized plane wave from the ambient, which makes 
an angle  $\theta_0$ with  the normal to the first 
interface and  has amplitude  $E_{a}^{(+)}$. 
The electric field is either in the plane of  incidence 
($p$ polarization) or perpendicular to the plane of 
incidence ($s$ polarization). We consider as well 
another plane wave of the same frequency and  
polarization, and with amplitude $E_{s}^{(-)}$, 
incident from the substrate at the same angle 
$\theta_0$~\cite{Snell}.

As a result of multiple reflections in all the interfaces,
we have a backward-traveling plane wave in the 
ambient, denoted $E_{a}^{(-)}$, and a  
forward-traveling plane wave in the substrate, 
denoted $E_{s}^{(+)}$. It is useful to treat the 
field amplitudes as the vector 
\begin{equation}
\label{Evec}
\mathbf{E} = 
\left ( \begin{array}{c}
E^{(+)} \\ 
E^{(-)} 
\end{array}
\right ) ,
\end{equation}
which applies to both ambient and substrate
media. Then, the amplitudes of the fields at
each side of the multilayer are related by a 
$2 \times 2$ complex matrix $\mathsf{M}_{as}$, 
we shall call a multilayer transfer matrix~\cite{OH00}, 
in the form 
\begin{equation}
\label{M1}
\mathbf{E}_a =  
\mathsf{M}_{as} 
\mathbf{E}_s ,
\end{equation}
where, in our case, $\mathsf{M}_{as}$ can be
expressed as
\begin{equation}
\label{Mlossless}
\mathsf{M}_{as} =
\left [
\begin{array}{cc}
1/T_{as} & R _{as}^\ast/T_{as}^\ast \\ 
R_{as}/T_{as} & 1/T_{as}^\ast
\end{array}
\right ]  
\equiv
\left [
\begin{array}{cc}
\alpha & \beta \\ 
 \beta^\ast & \alpha^\ast
\end{array}
\right ]  ,
\end{equation}
with $\det \mathsf{M}_{as} = +1$, which 
shows that the set of lossless multilayer
matrices reduces to the group SU(1,1), whose
elements depend on three independent real
parameters. In this matrix $R_{as}$ 
and $T_{as}$ are, respectively,  the overall 
reflection and  transmission coefficients for a 
wave incident  from the ambient. 

For the case at hand,  the Iwasawa decomposition 
reads as~\cite{MO01b} 
\begin{equation}
\label{Iwa1}
\mathsf{M}_{as} = 
\mathsf{K} (\phi)  
\mathsf{A} (\xi) 
\mathsf{N}(\nu) ,
\end{equation}
where
\begin{eqnarray}
\label{Iwasa1}
\mathsf{K} (\phi) & = & 
\left [
\begin{array}{cc}
\exp (i\phi/2) & 0 \\ 
0 & \exp (-i\phi/2)
\end{array}
\right ]  , 
\nonumber \\
\mathsf{A} (\xi) & = & 
\left [
\begin{array}{cc}
\cosh (\xi/2) & i \sinh(\xi/2) \\ 
-i \sinh(\xi/2) & \cosh (\xi/2)
\end{array}
\right ] , \\
\mathsf{N} (\nu) & = & 
 \left [
\begin{array}{cc}
1 - i \nu/2& \nu/2 \\ 
\nu/2 & 1+ i \nu/2
\end{array}
\right ]  .
\nonumber
\end{eqnarray}
In other words,  $\mathsf{K}(\phi)$  belongs
to a  compact,  $\mathsf{A}(\xi)$ to an abelian,
and $\mathsf{N}(\nu)$ to a nilpotent subgroup,
respectively, and their physical meaning has
been discussed in Ref.~\cite{MO01b}.  
Furthermore, such a  decomposition is 
global and unique~\cite{HE78}. 

As we have said, we are often interested in the 
transformation properties of field quotients rather 
than the fields themselves.  Therefore, it seems 
natural to consider the complex numbers
\begin{equation}
\label{defz}
z   =  \frac {E^{(-)}}{E^{(+)}} ,
\end{equation}
for both ambient and substrate. From a geometrical 
viewpoint, Eq.~(\ref{M1}) defines a transformation 
of  the complex plane ${\mathbb{C}}$, mapping the 
point $z_s$ into the point $z_a$, according to
\begin{equation}
\label{accion}
z_a = \Phi [\mathsf{M}_{as} , z_s] = 
\frac{\beta^\ast +\alpha^\ast z_s} 
{\alpha + \beta z_s}  .
\end{equation}
Thus, the action of the multilayer can be seen  as 
a function  $z_a = f(z_s)$ that can be appropriately 
called the multilayer transfer function~\cite{AZ87,MO01c}.
The action of the inverse matrix $\mathsf{M}_{as}^{-1}$ is 
$z_s = \Phi [\mathsf{M}_{as}^{-1}, z _a]$.

We can define an action of the group SU(1,1) on the 
complex plane $\mathbb{C}$ by these bilinear 
transformations. The complex  plane appears then 
decomposed  in three regions that remain invariant 
under the action of the group: the unit disc, 
its boundary and the external region~\cite{MO01c}. 

The Iwasawa decomposition has an immediate 
translation in this geometrical framework, and one 
is led to treat separately the action of each one of 
the matrices appearing in this decomposition. To
this end, it is worth noting that the group SU(1,1) 
we are considering appears always as a group of 
transformations of the complex plane. The concept 
of orbit is especially appropriate for obtaining an intuitive 
picture of the corresponding action. We recall that, given 
a point $z$,  its orbit is the set of  points $z^\prime$
obtained from $z$ by the action of  all the elements of 
the group. In Fig.~1 we have plotted some orbits 
for each one of the subgroups of  matrices 
$\mathsf{K}(\phi)$, $\mathsf{A} (\xi)$, and 
$\mathsf{N} (\nu)$. For matrices $\mathsf{K}(\phi)$ 
the orbits are circumferences centered at the origin,
for $\mathsf{A} (\xi)$, they are arcs of circumference  
going from the point $ +i$ to the point $-i$ through $z$. 
Finally, for the matrices $\mathsf{N} (\nu)$ the orbits 
are  circumferences passing through the point $+ i$ and 
joining the points $z$ and $-z^\ast$.

To go beyond this geometrical picture of multilayers, let us 
introduce the following classification: a matrix is of class 
$K$ when $[\mathrm{Tr}( \mathsf{M}_{as})]^2 < 4$,
is of class $A$ when  $[\mathrm{Tr} ( \mathsf{M}_{as})]^2 > 4$, 
and finally is of  class $N$ when $[\mathrm{Tr}( \mathsf{M}_{as})]^2 
= 4$. To gain insight into this classification, let us also 
introduce the fixed points~\cite{BA47} of a transfer matrix as the 
points in the complex plane that are invariant under 
the action of $\mathsf{M}_{as}$; i.e.,
\begin{equation}
\label{C}
z = \Phi[\mathsf{M}_{as}, z ] ,
\end{equation} 
whose solutions are given by
\begin{equation}
z = \frac{-i \mathrm{Im}\ \alpha \pm
\sqrt{(\mathrm{Re} \ \alpha)^2 -1}}
{\beta} ,
\end{equation}
Since $\mathrm{Tr} (\mathsf{M}_{as}) =2 
\mathrm{Re} \ \alpha$, one easily check that the 
matrices of class $K$ have two fixed points, one 
inside and other outside the unit disc, both related
by an inversion; the matrices of class $A$ have 
two fixed points both on the boundary of the unit 
disc and, finally,  the matrices class $N$ have 
only one (double) fixed point on the boundary
of the unit disc.

Now the origin of the notation for the classes should be 
clear: if one consider the Iwasawa decomposition~(\ref{Iwa1}), 
one can see that the matrices  $\mathsf{K} (\phi)$  are of the 
class $K$ with the origin as the fixed point in the unit disc,  
matrices $\mathsf{A} (\xi)$ are of the class $A$ with fixed points 
$+ i$ and $-i$ and   matrices $\mathsf{N} (\nu)$ are of the class 
$N$ with  the double fixed point $+ i$. Of course, this is in
agreement with the orbits in Fig.~1.

To proceed further let us note that by taking the conjugate
of $\mathsf{M}_{as}$ with any matrix 
$\mathsf{C}\in $ SU(1,1) we obtain another
multilayer matrix; i.e.,
\begin{equation}
\label{conjC}
\widehat{\mathsf{M}}_{as} = \mathsf{C}  \
\mathsf{M}_{as} \ {\mathsf{C}}^{-1} ,
\end{equation}
such that $\mathrm{Tr} (\widehat{\mathsf{M}}_{as}) =
\mathrm{Tr} (\mathsf{M}_{as})$. The fixed points
of $\widehat{\mathsf{M}}_{as}$ are then the image
by $\mathsf{C}$ of the fixed points of $\mathsf{M}_{as}$.
For our classification viewpoint it is essential to remark 
that if a multilayer has  a transfer matrix in the class 
$K$, $A$, or $N$, one can always find  a family 
of matrices $\mathsf{C}$ such that the conjugate 
through $\mathsf{C}$ is of the form  
$\mathsf{K}(\phi)$, $\mathsf{A}(\xi)$ or 
$\mathsf{N}(\nu)$,  respectively. The explicit 
construction of this family of matrices is easy: 
it suffices to impose that $\mathsf{C}$ transforms 
the fixed points of $\mathsf{M}_{as}$ into the 
corresponding fixed points of $\mathsf{K}(\phi)$, 
$\mathsf{A}(\xi)$, or  $\mathsf{N}(\nu)$. For example, 
if $\mathsf{M}_{as}$  is in the class $K$ and its
fixed point inside the unit disc is $z_f$, one should have
\begin{equation}
\Phi[\mathsf{C} \mathsf{M}_{as} {\mathsf{C}}^{-1}, 0]
= \Phi[\mathsf{C} \mathsf{M}_{as}, z_f] = 
\Phi[\mathsf{C}, z_f] = 0 .
\end{equation}

In order to appreciate the physical meaning of this 
classification let us take another conjugate of 
Eq.~(\ref{conjC}) by  the  unitary matrix 
\begin{equation}
{\mathcal{U}} =
\frac{1}{\sqrt{2}}
\left [
\begin{array}{cc}
1 & i \\ 
i & 1
\end{array}
\right ]   .
\end{equation}
In consequence, we can rewrite it alternatively as
\begin{equation}
\bm{{\mathcal{E}}}_a =  
{\mathcal{M}}_{as} 
\bm{{\mathcal{E}}}_s ,
\end{equation}
where the new field vectors are defined as
\begin{equation}
\bm{{\mathcal{E}}} =  
\left ( \begin{array}{c}
{\mathcal{E}}^{(+)} \\ 
{\mathcal{E}}^{(-)} 
\end{array}
\right ) =
{\mathcal{U}} 
\left ( \begin{array}{c}
\widehat{E}^{(+)} \\ 
\widehat{E}^{(-)} 
\end{array}
\right ) =
{\mathcal{U}} \mathsf{C}
\left ( \begin{array}{c}
E^{(+)} \\ 
E^{(-)} 
\end{array}
\right ) ,
\end{equation}
and the transformed multilayer matrix reads as
\begin{equation}
\mathcal{M}_{as} =  
\mathcal{U} \ \widehat{\mathsf{M}}_{as} 
\ {\mathcal{U}}^{-1} =
\mathcal{U} \mathsf{C} \ \mathsf{M}_{as} \ 
{\mathsf{C}}^{-1} {\mathcal{U}}^{-1} .
\end{equation}
One can easily check that $\det{\mathcal{M}}_{as} = +1$ 
and all its elements are real numbers.  Therefore, 
${\mathcal{M}}_{as}$ belongs to the group SL(2,$\mathbb{R}$)
that underlies the structure of the celebrated $ABCD$ law
in first-order optics~\cite{geom1,geom2,geom3}. 

By transforming by ${\mathcal{U}}$ the Iwasawa 
decomposition (\ref{Iwa1}), we get  the corresponding 
one for SL(2,$\mathbb{R}$), which has been 
previously worked out~\cite{Iwasl2r}:
\begin{equation}
\label{Iwa2}
{\mathcal{M}}_{as} = 
{\mathcal{K}}(\phi)  
{\mathcal{A}}(\xi) 
{\mathcal{N}}(\nu) ,
\end{equation}
where
\begin{eqnarray}
{\mathcal{K}}(\phi)  & = & 
\left [
\begin{array}{cc}
\cos(\phi/2) &  \sin(\phi/2) \\ 
- \sin(\phi/2)  & \cos(\phi/2) 
\end{array}
\right ]  , 
\nonumber \\
{\mathcal{A}}(\xi) & = & 
\left [
\begin{array}{cc}
\exp(\xi/2) & 0 \\ 
0 & \exp (-\xi/2)
\end{array}
\right ] , \\
{\mathcal{N}}(\nu) & = & 
 \left [
\begin{array}{cc}
1 & 0 \\ 
 \nu & 1
\end{array}
\right ]  .
\nonumber
\end{eqnarray}
The physical action of these matrices is clear.
Let us consider by the moment all of them as $ABCD$ 
matrices in geometrical optics that apply to position 
$\mathbf{x}$ and momentum $\mathbf{p}$ 
(direction) coordinates of a ray in a transverse plane.
These are the natural phase-space variables
of ray optics. Then $\mathcal{K} (\phi)$ would 
represent a rotation in  these variables,  $\mathcal{A} (\xi)$ 
a magnifier that scales $\mathbf{x}$ up to the factor 
$m = \exp(\xi/2)$ and $\mathbf{p}$ down by the same 
factor, and $\mathcal{N} (\nu)$  the action of a lens 
of power $\nu$~\cite{geom3}.   

In the multilayer picture, ${\mathcal{E}}^{(+)}$ 
can be seen as the corresponding  $\mathbf{x}$,
while ${\mathcal{E}}^{(-)}$ can be seen as
the corresponding $\mathbf{p}$. Then, the key
result of this Letter is that when the multilayer
transfer matrix has $[\mathrm{Tr} 
({\mathsf{M}}_{as}) ]^2$ lesser, greater or equal
to 4 one can find in a direct way a family of matrices
that gives a new vector basis such that the action
of the multilayer, when viewed in such a basis,
is exclusively rotationlike, or magnifierlike,  or lenslike.

We expect that the formalism presented here could 
provide a general tool for analyzing and classifying the 
multilayer performance in an elegant and concise way.

\newpage

\begin{figure}
\caption{Plot of several orbits in the unit disc 
of the elements of the Iwasawa decomposition 
$\mathsf{K}(\phi)$,  $\mathsf{A}(\xi)$,  and 
$\mathsf{N}(\nu)$ for the group of multilayer 
transfer matrices.}
\end{figure}


\begin{thebibliography}{99}
\bibitem{Snell}
When ambient ($0$) and substrate ($m+1$) 
media  are different, the angles  $\theta _{0}$ 
and  $\theta _{m+1}$ are conected by Snell law 
$n_0\sin \theta_0=  n_{m+1}\sin \theta_{m+1}$, 
where $n_j$ denotes the refractive index of the 
$j$th medium.
\end{thebibliography}
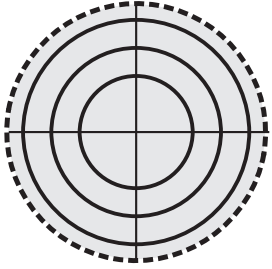
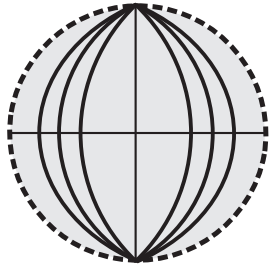
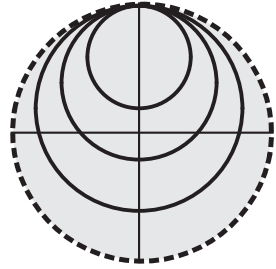
\end{document}